\documentclass{PoS}

\title{Hadron production in hot and dense nuclear matter}

\ShortTitle{Hadron production in hot and dense nuclear matter}

\author{\speaker{A. Lavagno}\\
        Dipartimento di Fisica, Politecnico di Torino and INFN,
        Sezione di Torino, Italy\\
        E-mail: \email{andrea.lavagno@polito.it}}

\abstract{We study the hadron production at finite values of temperature and baryon density by means of an effective relativistic mean-field model with the inclusion of the full octet of baryons, the $\Delta$-isobar degrees of freedom and the lightest pseudoscalar and vector mesons. These last particles are considered in the so-called one-body contribution, taking into account of an effective chemical potential and mass depending on the self-consistent interaction between baryons. The analysis is performed by requiring the Gibbs conditions on the global conservation of baryon number, electric charge fraction and zero net strangeness. In this context, we study the behavior of different particle-antiparticle ratios and strangeness production.}

\FullConference{The 2011 Europhysics Conference on High Energy Physics, EPS-HEP 2011,\\
		July 21-27, 2011\\
		Grenoble, Rh\^one-Alpes, France}

\begin{document}

We study the hadronic EOS by requiring the global conservation of baryon number, electric charge fraction and zero net strangeness in the range of finite temperature and density. The study is performed by means of an effective relativistic mean field (RMF) model with the inclusion of the octet of the lightest baryons, the $\Delta$-isobar degrees of freedom and the lightest pseudoscalar and vector mesons \cite{serot-wal,greiner97,glen_prl1991}. These last particles have been considered in the so-called one-body contribution, taking into account their effective chemical potentials depending on the self-consistent interaction between baryons \cite{prc2010}. In analogy to the effective meson chemical potentials, the effective meson masses are expressed as a difference of the effective baryon masses respecting the strong interaction and the main processes of the meson production/absorption involving different baryons. Such assumption implies a mechanics in which the vacuum meson masses are reduced (enhanced) if the ratio between the $\sigma$-meson field coupling with the heavier baryon ($\Delta$ or hyperon particles involved in the meson production/absorption) and the nucleon one, is greater (lower) than one. A variation of the effective meson masses in-medium simulates, in our simple scheme, the relevance of meson-meson and meson-baryon self-interaction in the nuclear medium at finite temperatures and baryon densities.

The chemical potential of particle of index $i$, $\mu_i$, can be expressed in term of
the three independent chemical potentials: $\mu_i=b_i\, \mu_B+c_i\,\mu_C+s_i\,\mu_S$,
where $b_i$, $c_i$ and $s_i$ are, respectively, the baryon, the
electric charge and the strangeness quantum numbers of the $i$-th
hadronic species.
At low baryon density and high temperature, the contribution of the lightest pseudoscalar and
vector mesons to the total thermodynamical potential becomes very
relevant. From a phenomenological point of view, we can take into
account of these contributions by incorporating such mesons by
adding to the thermodynamical potential their one-body
contribution, i.e. the contribution of an ideal Bose gas with an
effective chemical potential.
Following Ref. \cite{prc2010}, the values of the meson effective chemical potentials are fixed from the "bare" chemical potentials and writing them in terms of the
corresponding baryon effective chemical potentials, respecting the
strong interaction. For example, for pions (and rho mesons) we
have that $\mu_{\pi^+}=\mu_{\rho^+}=\mu_C\equiv\mu_p-\mu_n$ and
its effective chemical potential can be written as
\begin{equation}
\mu_{\pi^+}^*=\mu_{\rho^+}^*\equiv\mu_p^*-\mu_n^* \, .
\end{equation}
For the other mesons, we have
\begin{eqnarray}
&&\mu_{K^+}^*=\mu^*_{K^{*+}}\equiv\mu_p^*-\mu_{\Lambda(\Sigma^0)}^* \, ,\\
&&\mu_{K^0}^*=\mu^*_{K^{*0}}\equiv\mu_n^*-\mu_{\Lambda(\Sigma^0)}^*
\, ,
\end{eqnarray}
while the others non-strange neutral mesons have a vanishing
chemical potential. Thus, the effective meson chemical potentials
are coupled with the meson fields related to the interaction
between baryons. This assumption
represents a crucial feature in the EOS at finite density and
temperature and can be seen somehow in analogy with the hadron
resonance gas within the excluded-volume approximation. There the
hadronic system is still regarded as an ideal gas but in the
volume reduced by the volume occupied by constituents (usually
assumed as a phenomenological model parameter), here we have a
(quasi free) mesons gas but with an effective chemical potential
which contains the self-consistent interaction of the meson
fields.

The numerical evaluation of the above thermodynamical quantities
can be performed if the meson-nucleon, meson-$\Delta$ and
meson-hyperon coupling constants are known. Concerning the
meson-nucleon coupling constants they are determined to reproduce
properties of equilibrium nuclear matter such as the saturation
densities, the binding energy, the symmetric energy coefficient,
the compression modulus and the effective Dirac mass at
saturation. The set marked GM3 is from Ref. \cite{glen_prl1991}.
The implementation of hyperon degrees of freedom comes from
determination of the corresponding meson-hyperon coupling
constants that have been fitted to hypernuclear properties \cite{gal}. We refer to \cite{prc2010} for a detailed discussion on the dependence of the EOS from different $\Delta$-couplings).

\begin{figure}[htb]
\begin{center}
\resizebox{0.48\textwidth}{!}{%
\includegraphics{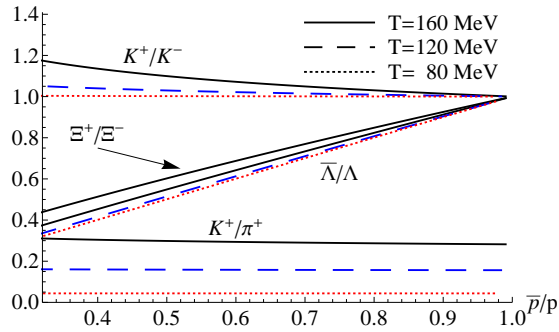}}
\caption{Particle-antiparticle and $K^+/\pi^+$ ratios as a
function of the $\overline{p}/p$ ratio for different temperatures.
The $\Delta$ coupling ratios are fixed to $r_s=r_v=1$. The ratios
of $\Xi^+/\Xi^-$ at $T=80$, $120$ MeV are not reported because
they are very strictly to the $\overline{\Lambda}/\Lambda$ ones. }
\label{antipar}
\end{center}
\end{figure}

In Fig. \ref{antipar}, we report the results of various
particle-antiparticle ratios and $K^+/\pi^+$ ratio as a function
of the $\overline{p}/p$ ratio for different values of temperature.
We can observe good agreement with the results obtained in the framework of
statistical-thermal models and with experimental SPS
and RHIC data \cite{star_prc09}.

In conclusion, we have studied an effective relativistic mean field model with the inclusion of the full octet of baryons, the $\Delta$-isobars and the lightest pseudoscalar and vector mesons, by requiring, in the range of finite density and temperature, the global conservation of baryon number, electric charge fraction and zero net strangeness.
The meson degrees of freedom have been incorporated in the EOS as a quasi-particle Bose gas with an effective meson chemical potential $\mu^*$ expressed in terms of meson fields, responsible for the self-consistent mean field interaction. The introduced effective EOS is principally devoted to a regime of finite and intermediate values of baryon density and temperature reachable in the future
compressed baryonic matter experiments.

\end{document}